\documentclass[11pt]{article} 
\usepackage[utf8]{inputenc}
\usepackage{multirow}
\usepackage{amsfonts}
\usepackage{amsmath}
\usepackage{amssymb}
\usepackage{amsthm}
\usepackage{caption}
\usepackage[dvipsnames]{xcolor}
    \definecolor{darkgreen}{rgb}{0,0.5,0}
    \definecolor{darkblue}{rgb}{0,0,0.6}
    \definecolor{purple}{rgb}{0.4,.2,0.7}
\usepackage[margin = 2.5cm]{geometry}
\usepackage{graphicx}
\usepackage[hyperfootnotes = false, colorlinks = true, linkcolor = darkblue, citecolor = purple]{hyperref}
\usepackage{subcaption}

\newcommand{\be}{\begin{equation}}
\newcommand{\ee}{\end{equation}}
\newcommand{\bea}{\begin{eqnarray}}
\newcommand{\eea}{\end{eqnarray}}
\def\la{\label}
\def\nref#1{(\ref{#1})}
\def\half{{1 \over 2 }}

	\newcommand{\bes}{\begin{equation} \begin{split} }	
	\newcommand{\ees}{\end{split} \end{equation} }

	\usepackage{physics}

\begin{document}

\thispagestyle{empty}
\begin{center}
    ~\vspace{5mm}

  {\LARGE \bf {Comments on the no boundary wavefunction and slow roll inflation  \\}}



   \vspace{0.5in}
     
   {\bf     Juan Maldacena$^1$ }

    \vspace{0.5in}
 
   ~
   \\
   $^1$Institute for Advanced Study,  Princeton, NJ 08540, USA

    \vspace{0.5in}

    \vspace{0.5in}
    

\end{center}

\vspace{0.5in}

\begin{abstract}
 We review aspects of the Hartle-Hawking no boundary geometry in the context of slow roll inflation. 
 We give an analytic approximation to the geometry and we explain the rationale for the proposal. 
 We also explain why it gives a prediction for the curvature of the universe that is in disagreement with observations and give a quick review of proposed ways to resolve that disagreement. 
  
 \end{abstract}

\vspace{1in}

\vspace{1in}
{\it ~~~~~~~~~~~~~~~~~~~~~~Presented at the James Hartle memorial conference} 

\pagebreak

\setcounter{tocdepth}{3}
{\hypersetup{linkcolor=black}\tableofcontents}

 \section{Introduction} 
 
 The no boundary wavefunction is a very interesting proposal for the wavefunction of the universe \cite{Hartle:1983ai}. 
 We consider the wavefunction of the universe as a function of the three dimensional spatial geometry   and the values of various fields living on that geometry. We denote these by $h_{ij}$ and $\phi$.   The proposal involves  a (generically) complex four dimensional  geometry which ends  at the given spatial 3-geometry and has no further boundary \cite{Halliwell:2018ejl}. In particular, it has no boundary into the past. Given this no-boundary geometry we compute the wavefunction via 
 \be 
 \Psi[ h_{ij}, \phi] \propto \exp\left( i I[g_{\mu \nu},\phi]  \right) \la{NBp}
 \ee 
 where $I$ is the    classical action evaluated on  the no boundary four dimensional geometry given by the metric $g_{\mu \nu}$ and fields $\phi$.\footnote{ The proposal can be trivially extended to other spacetime dimensions.} 
 
 This is a very elegant and theoretically compelling proposal but it unfortunately seems to give the wrong answer when we try to apply it to our universe. In the context of slow roll inflation,  it gives a result proportional to 
 \be 
 |\Psi|^2 \propto \exp( {24 \pi^2 } { M_p^4 \over V(\phi_*)} )  ~,~~~~~~~ M_p^{-2} = 8 \pi G_N
  \la{HHInt}
 \ee 
 to create a universe that starts inflating when the inflaton takes the value $\phi_*$,  where $V(\phi)$ is the potential for the scalar field. So it gives a very large weight to values of $\phi$ where the potential is very small.    Taking the potential to its present value we would get an empty universe with a probability that dominates  by an amount of order $\exp( 10^{120} )$, the exponential of the cosmological constant problem!. 
 
 Here we will  review this proposal and describe how it applies to slow roll inflation \cite{Hartle:2007gi,Janssen:2020pii}. In this context, it gives a prediction for the probability that the spatial sections of the universe have positive scalar curvature. It predicts that the curvature should be much larger than what is observed. 
 
 We   emphasize that this proposal is very natural in the inflationary context and it represents a small generalization of the highly successful inflationary prediction for primordial curvature perturbations \cite{Mukhanov:1981xt,Starobinsky:1982ee,Guth:1982ec,Hawking:1982cz,Bardeen:1983qw}.
      One way to phrase it is that it gives curvature modes with distance scales shorter than the size of the universe in good agreement with experiment \cite{bennett2013nineyear,Planck:2018jri}\footnote{The observed amplitudes for low $\ell$ modes are somewhat smaller than expected \cite{bennett2013nineyear,Planck:2018jri}, but this seems completely unimportant compared to the problem with the constant curvature mode. },  while it produces a completely wrong answer for the overall constant curvature mode!. 
 
 This problem is certainly not new \cite{Vilenkin:1987kf}, it is well known to specialists,  and there have been various proposals for its resolution, which we will attempt to partially review in section \ref{Solutions}. For another recent review of the no boundary wavefunction see \cite{Lehners:2023yrj}. 

 We will  take the opportunity to describe  analytical approximations to the no boundary geometry in the slow roll approximation, geometries that were previously described mainly numerically. As a by product we explain why the KSW criterion \cite{Kontsevich:2021dmb,Witten:2021nzp} constrains $r $ \cite{Hertog:2023vot}.  We will also review the rationale behind the no boundary proposal to explain why it is a natural extension of the computations that we normally do for small perturbations in inflation.

 \section{Short summary of the no boundary geometry in the slow roll inflation situation} 
 
 In this section,  we present the physical setup and summarize the result. In the later sections we will discuss it in more detail.   
 
\begin{figure}[t]
   \begin{center}
   \includegraphics[scale=.4]{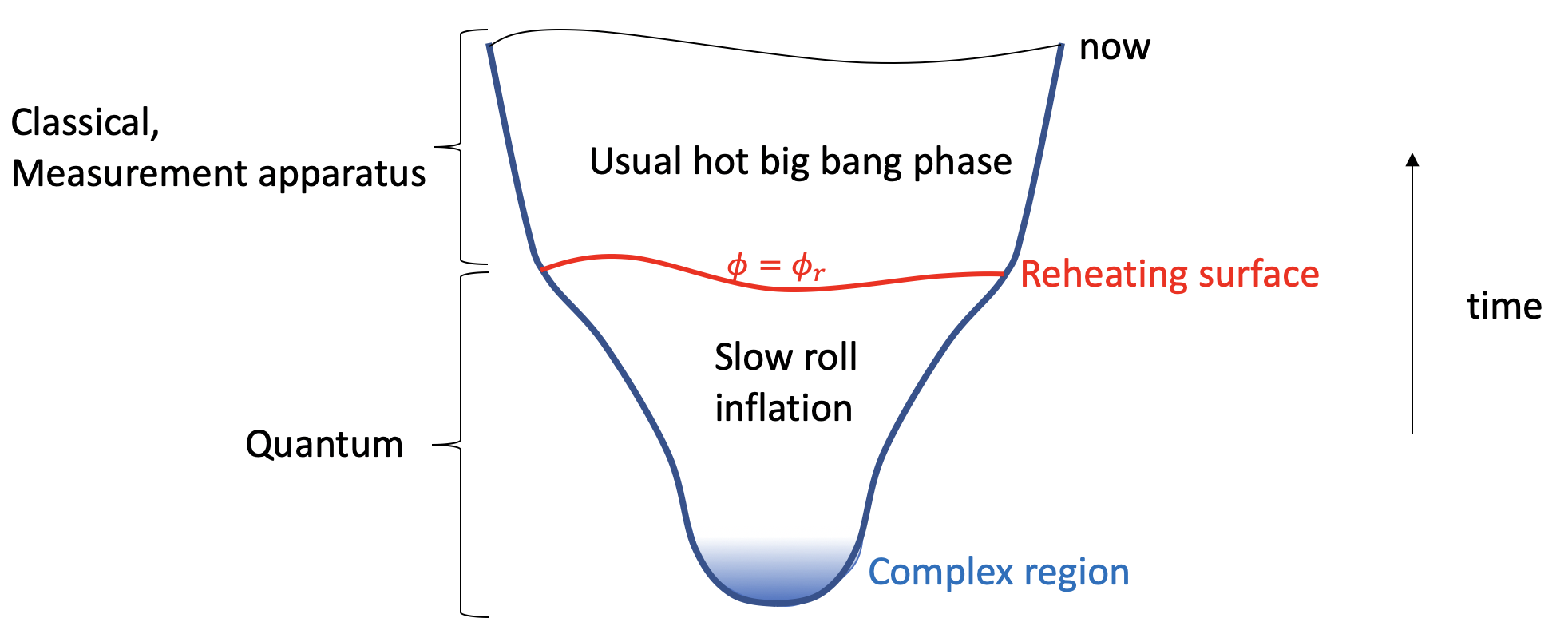}
    \end{center}
    \caption{Cartoon for the evolution of the universe. We have a period of inflation followed by the standard hot big bang picture. We treat the inflationary period using quantum mechanics up to the reheating surface. The universe that follows is treated classically and as a measurement apparatus that measures the shape of the reheating surface.    }
   \label{Reheating}
\end{figure}

 Let us consider a cosmic history containing a period of single field slow roll inflation, with inflation ending, reheating the universe and giving rise to the universe we see. 
 We assume that the end of inflation and reheating is relatively quick. We consider the problem of predicting the shape of the spatial slice at reheating, or equivalently, near the end of the inflationary period. This is equivalent to saying that we want to predict the geometry of the spatial slices when the scalar field takes a value $\phi_r$ close to its reheating value, but before reheating, see figure \nref{Reheating}. We are  picking a slice with a constant inflaton value. Alternatively, we can say that we are viewing  the inflaton as the clock.
 
 For the purposes of this paper,  we treat the inflationary period quantum mechanically, using the usual semiclassical path integral for gravity, and the later big bang phase is treated classically as a measurement apparatus for the earlier universe. This is a reasonable approximation for the computation of the small curvature fluctuations,  and we will also use it to compute the overall curvature of the universe. 
  We take this point of view so that we do not have to worry about the potential taking the present value in \nref{HHInt}. The reader is justified in thinking that this approximation might not be appropriate for the problem of computing the overall spatial curvature of the universe. We will   work in this restricted context in this paper, since it leads to a clearer discussion and it is instructive to discuss first this case.  We leave  further discussion to section \ref{Solutions}.

  Inflation classically predicts that this slice is essentially flat. However, quantum effects during inflation predict that the slice should have small curvature fluctuations. 
 
Parametrizing the spatial metric near the end of the inflationary period as  
\be \la{ScFl}
 ds^2 = a_r^2  e^{ 2 \zeta(x) } d\vec x^{\,2} 
 \ee 
  where $a_r$   is the scale factor for a homogeneous solution and $\zeta(x)$ is a change in the scale factor of the metric\footnote{There are also tensor fluctuations, but we will neglect them for our discussion.}. We are considering fluctuations over scales much larger than the size of the horizon at the time where inflation ends, so that $\zeta $ is essentially constant in time. 
  
 The standard inflationary computation  \cite{Mukhanov:1981xt,Starobinsky:1982ee,Guth:1982ec,Hawking:1982cz,Bardeen:1983qw} gives us a probability  that has the rough form 
 \be \la{Sfc}
  |\Psi(\zeta)|^2 \sim \exp\left( -  \int { d^ 3 k \over (2\pi)^3 } {2 \epsilon_*   \over H_*^2 }  {   k^3 } \zeta_{k } \zeta_{-k } \right) 
  \ee 
  in fourier space.  Here we defined 
  \be 
 \epsilon = { {V'}^2 \over 2 V^2 } ~,~~~~~~~~~~ 8 \pi G_N =M_p^{-2} =1 , ~~~~~~~ 3 H^2 \sim V 
 \ee 
 The star in \nref{Sfc} means that we are evaluating $H$ and $\epsilon$ at the time where the mode with momentum $k$ crosses the horizon 
 \be \la{HorCro}
    { k \over a(\phi_*) } \sim  H(\phi_*)  ~~~~~~~~~\to ~~~~~~~\phi_*(k)
    \ee

 The probability distribution  \nref{Sfc}  is  in broad agreement with observations. 
 Note that we can say that we are computing the curvature fluctuations of the spatial slice  since, $R^{(3)} \propto - { 1 \over a_r^2 } \nabla^2 \zeta \sim { k^2\over a_r^2 } \zeta$. 
 
  \begin{figure}[t]
   \begin{center}
   \includegraphics[scale=.4]{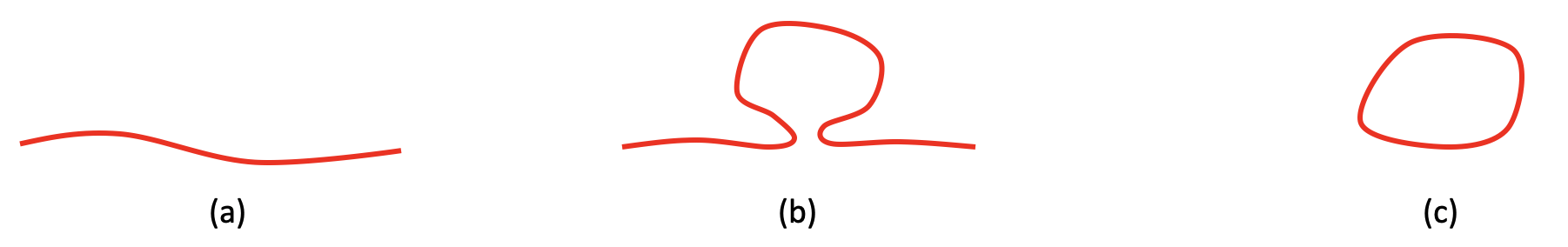}
    \end{center}
    \caption{We can use complex classical solutions to compute the wavefunctions for universes of various shapes. (a) The usual case of almost flat slices with small fluctuations. (b) A situation with a large fluctuation. (c) A closed universe. }
   \label{Shapes}
\end{figure}
 
 The no boundary wavefunction allows us to consider   large curvature fluctuations, see figure \ref{Shapes}b. In principle, we can compute such probabilities using complex classical solutions. 
 A simple case corresponds to a spherical shape of the form 
 \be 
 ds^2 =  a_r^2   d\Omega_3^2
  \ee 
  or curvature of order $1/a_r^2$. 
  We consider a no boundary solution which has $\phi = \phi_r$ at the boundary. Then in the region where the sphere shrinks to zero the value of $\phi$ will be of order $\phi_*$.    As we will review below, the no boundary answer for the probability goes as 
  \be \la{ProbC}
  |\Psi|^2 \propto \exp\left( { 8 \pi^2 } { 3 \over V(\phi_*) }  \right) 
  \ee 
  where, in analogy to \nref{HorCro}, the value of $\phi_*$ depends on $a_r$ and $\phi_r$  and is set by the ``horizon crossing" value 
  \be 
   { 1 \over   a_f(\phi_*) } = H(\phi_*) ~,~~~~~~~~~~{\rm with } ~~~~~~a_f(\phi_r) = a_r 
   \la{PhiS}
   \ee 
   and we assumed we have a solution $a_f(\phi)$ for the usual inflationary solution with flat slices. 
   
   We see that a smaller value of $a_r$ means that   horizon crossing will occur later during the inflationary trajectory,  which means that $V(\phi_*)$ will be smaller and the probability \nref{ProbC} will be larger. 
   
  The solution becomes complex around the time corresponding to $\phi_*$, as we will see in section \ref{HHSlow}.     Therefore we can think of $\phi_*$ in \nref{PhiS} as   the start of inflation and we find that the physical size of the sphere at the reheating surface is
  \be
   a_r \propto { 1 \over H(\phi_*) } \exp({\cal N }  ) ~~~~~~~~,~~~~~~{\cal N }  \sim -\int_{\phi_*}^{\phi_r} d\phi { V \over V' } 
   \ee 
    The result \nref{ProbC} implies that if we change the size of the sphere as 
    \be 
     a_r \to \tilde a_r = a_r e^{\Delta {\cal N }  }
     \ee 
        we will get a change in probability of the form 
    \be 
    { |\tilde \Psi |^2    \over |\Psi|^2 } \propto \exp\left( -8 \pi^2 { 3 \over V_*} { V'^2_* \over V_*^2 }    \Delta  {\cal N }   \right) \propto \exp\left( - { 2 \over A_s}   \Delta  {\cal N } \right) = \exp\left( - { 2 \over A_s}  \log\left[{ \tilde a_r \over a_r } \right] \right) ~,~~~~~~~{ 2 \over A_s} \sim 10^9 \la{Shife}
  \ee 
  where we noted that the combination of parameters is the same as the one setting the square of the amplitude of scalar fluctuations \nref{Sfc} and we quoted  its value from \cite{Planck:2018jri} to emphasize that it is a known large value.  
  
  This implies that we gain in probability by  reducing the number of efolds, or the size of the sphere at the reheating surface. 
  
  In our universe,  we know that the size of the sphere, if there was one,  is now larger than the observable universe, since $|\Omega_k| = { 1 \over a_{\rm now}^2 H_{\rm now}^2} \lesssim 0.003$ \cite{Planck:2018jri}.  The result \nref{Shife} implies a huge probability pressure to reduce this size. We could wonder, for example, why we could not have say $\Omega_k \sim .1$ rather than the smaller experimentally measured value. This would certainly not make a significant change in our present universe, so that it seems unlikely to be forbidden by an anthropic argument.

  
  We emphasize that the overall curvature of the universe is essentially the same type of observable that we observe when we look at the small  curvature fluctuations. For the small curvature fluctuations, or $\ell \geq 2$ modes,  we use quantum mechanics to predict their value. Why should we not  be able to use quantum mechanics to predict the overall curvature of the universe too?. 

 In the rest of this article we give more details on the no boundary geometry used to derive these results. 
  
  We will discuss possible ways around this uncomfortable result in section  \ref{Solutions}.

  \section{Preliminaries: computing wavefunctions in flat space and de Sitter} 
  
   In this section,  we review how to compute vacuum wavefunctions using path integrals over imaginary time. The purpose is to review a familiar computation so that the no boundary proposal becomes more natural. 
   
   Let us first start with a simple case, a harmonic oscillator with action 
   \be 
   I = \int dt \half ( \dot x^2 - x^2 ) 
   \ee 
   As is well known, we can get the wavefunction of the ground state by evolving over imaginary time,  
   setting  $t = i t_E$  and going from   $t_E = + \infty$ to $t_E =0$.   
   The classical solution obeying $x(0) = x_0$ is $x(t_E)  = x_0 e^{ - t_E}$.
   Evaluating the classical action we get 
   \bea 
   i I &= &    - \int^{\infty}_0 dt_E e^{ - 2 t_E} x^2 =  - \half x_0^2 ~,~~~~\to~~~
   \Psi(x_0) \propto  \exp( i I) = \exp( -\half x^2_0 ) 
   \eea 
   which is   the usual answer, of course.
   
\begin{figure}[t]
   \begin{center}
   \includegraphics[scale=.4]{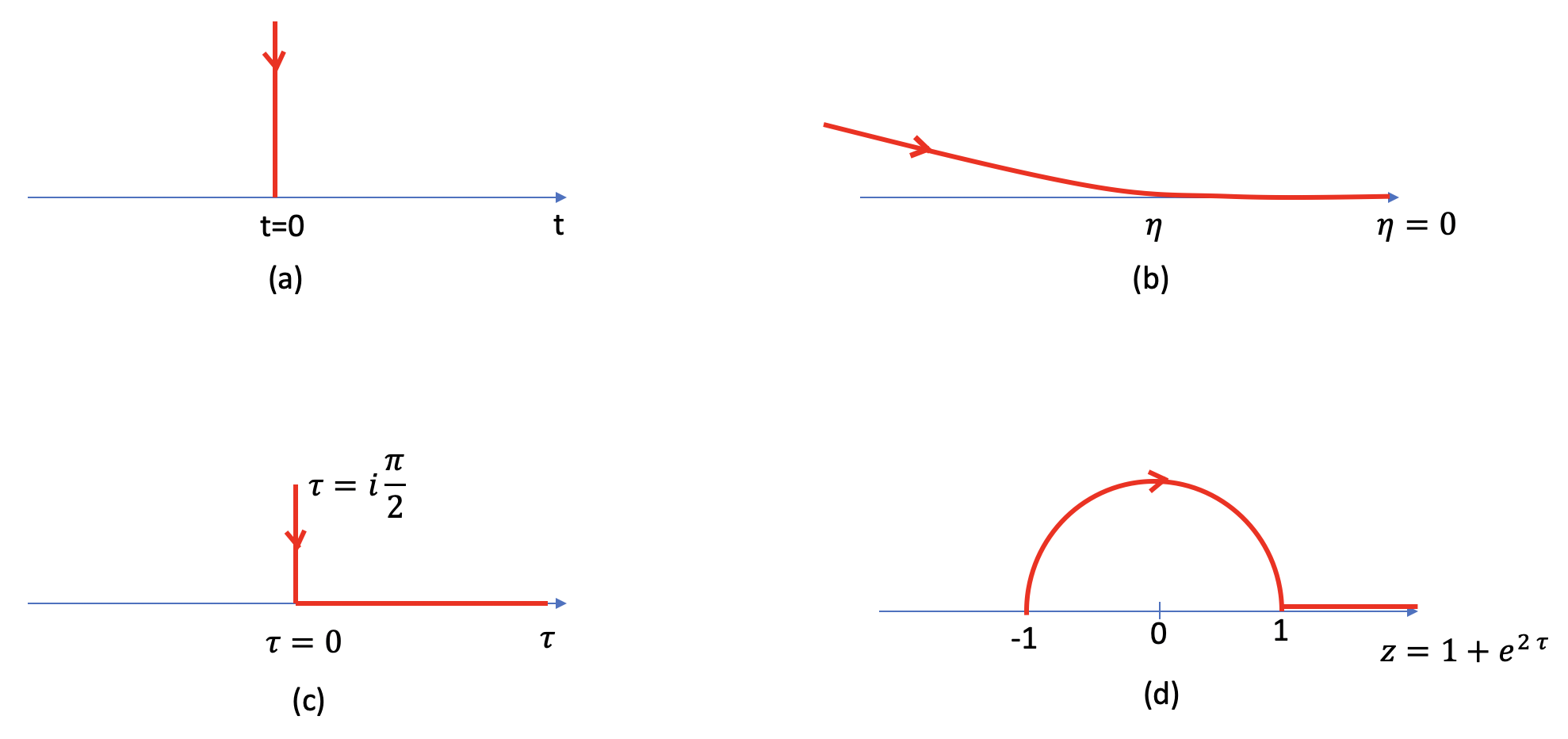}
    \end{center}
    \caption{Contours in the complex plane for various wavefunction computations. (a) For a harmonic oscillator or quantum fields in flat space a contour in imaginary time from infinity to $t=0$ gives us the wavefunction at $t=0$. (b) For a quantum field in de-Sitter with flat slicing \nref{Fsli} we start from imaginary values of $\eta$ and we evolve to real $\eta$ up to $\eta=0$ to find the wavefunction in the asymptotic future. (c) In de-Sitter in global slicking \nref{dSGlob} we start from $t= i { \pi \over 2}$ and then we go to $t=0$ and continue to real times. In (d) we see the argument of the function in \nref{HypSph} as we follow the contour in (c).   }
   \label{Contours}
\end{figure}

   In quantum field theory we can do the same. For a free theory it is the same as above since a free theory is a collection of harmonic oscillators. For an interacting theory we can also do this and obtain the wavefunction by evolving in Euclidean time. We note that the contour comes in from the positive imaginary time direction, see figure \ref{Contours}.

   \subsection{Scalar field in de Sitter in the flat slicing } 
   
   We can do the same for  a field in de-Sitter space, written in flat slices as 
   \be \la{Fsli}
   ds^2 =  { - d\eta^2 + d\vec x^2 \over \eta^2 } 
   \ee 
   For simplicity, let us say that we have a massless field with action 
   \be 
   I = \int \half { 1 \over \eta^2 } [  (\partial_\eta \phi)^2 - (\vec \partial_x\phi)^2 ] 
   \ee 
   Then the classical solution for a mode with momentum $k$ is 
   \be \la{flaWF}
   \phi_{\vec k}(\eta) = \hat \phi_{\vec k}  e^{ i \vec k . \vec x } e^{ i k \eta }( 1- i k \eta ) ~,~~~~~~k = |\vec k| 
   \ee 
   where $\hat \phi_{\vec k}$ is the fourier transform of the  field in the far future, $\eta \to 0^-$. Here we picked the solution which decays when Im$(\eta) \to + \infty$, which is the condition that results from thinking that the contour comes from the positive imaginary direction, see figure \ref{Contours}b.  Decomposing the field in Fourier modes as usual we find that 
   \be \la{Wfs}
   \Psi \sim \exp( i I ) \sim \exp\left[    \int { d^3 k \over (2 \pi )^3 } \left( i {k^2 \over 2\eta_c} - { k^3 \over 2} \right)   \hat \phi_{\vec k} \hat \phi_{- \vec k}  \right]
  \ee 
  For a real field $\hat \phi(x)$, we have that $\hat \phi_{- \vec k} = { \hat \phi}_{\vec k}^* $. The phase\footnote{ The phase factor   is a local term of the form $\int (\nabla \hat \phi)^2 $.} in  \nref{Wfs} drops out when we compute $|\Psi|^2$ and we get a result closely related to \nref{Sfc}.    
 
  Note that the solution \nref{flaWF} has the small $\eta$ expansion 
  \be \la{Soln}
  \phi_{\vec k}(\eta)  = \hat \phi_{\vec k} \left[ 1 + \half {\vec k}^{\,2} \eta^2 + i {1 \over 3} |\vec k|^3 \eta^3 + \cdots \right] 
  \ee 
  we see that in Fourier space, the first imaginary part in the solution arises at order $\eta^3$. 
  Notice that the wave equation near $\eta =0$ can be expanded in powers of $\eta$ and there are two independent solutions, one starts with $\eta^0$ and another starting with $\eta^3$. The one starting with $\eta^0$ is fixed by the boundary conditions at $\eta=0$ and includes the first two terms in \nref{Soln}. The one starting with $\eta^3$ is determined once we impose the boundary conditions at $\eta \to ( -1 + i \epsilon) \infty $. Its complex coefficient is related to the fact that the boundary conditions involve a complex deformation. A point we want to emphasize is that  the solution that is relevant for computing the wavefunction is complex. Just to be very clear about this point, note that the boundary conditions are real in position space,  $\hat \phi(x)$ is real. In order to have a real expression for $\phi(\eta , \vec x)$ we would need that $\phi_{\vec k}(\eta)^* = \phi_{-\vec k}(\eta)$. This is spoiled by the third term in \nref{Soln}. 
 The fact that the solution is complex is responsible for 
  the real part in the exponent in \nref{Wfs}, which is crucial for obtaining a reasonable normalizable wavefunction.  
 
 The inflationary result \nref{Sfc} arises   from \nref{Wfs} after restoring the Hubble constant dependence in \nref{Wfs}, which gives a factor of $H^{-2} $ in the exponent and connecting the scalar field to the curvature perturbation by $ \phi= { \dot \phi \over H } \zeta $ (see \cite{Maldacena:2002vr} for a more thorough review). 
  
  \subsection{Scalar field in de Sitter in the global slicing } 
  
   We can also consider de-Sitter in global slices 
  \be \la{dSGlob}
  ds^2 = - d\tau^2 + \cosh^2 \tau d\Omega_3^2 
  \ee 
   This metric can be continued   to $\tau = i \theta$ 
  \be 
  ds^2 = d\theta^2 + \cos^2 \theta d\Omega_3^2 
  \ee 
  
  The contour that selects the vacuum starts at $\theta = \pi/2$ or $\tau = i \pi/2$, it runs then down to real $\tau $ and then  goes to large real $\tau$, see figure \ref{Contours}c.  
  Again,    decomposing the massless field in angular momentum modes, writing   the wave equation, solving  it,  and selecting the solution that is regular at $\tau = i \pi/2$ we get 
  \be \la{HypSph}
  \phi \propto    e^{ 3 \tau } (1 + e^{ 2 \tau } )^\ell ~_2F_1\left( { 3 \over 2 } + \ell , 3 + \ell , 3 + 2 \ell ; 1 + e^{ 2 \tau } \right) 
  \ee 
  This obeys the right boundary condition at $e^{ 2 \tau } =-1$. 
  The real time solution should be obtained from this one by applying the   transformation rule  for hypergeometric functions. In doing so, it is important to note that along our contour the last argument of \nref{HypSph} describes a semicircle around one in the positive part of the complex plane, see figure \ref{Contours}d.  The result is proportional to
  \bea \la{SolLt}
 \phi & \propto &     (1 + e^{ - 2 \tau } )^\ell  \left[  ~_2F_1( {3 \over 2 } + \ell, \ell, -\half;  - e^{ -2 \tau }     )  - 
   i { 8 \over 3 }  \ell (\ell+1)(\ell +2)  e^{ - 3 \tau } ~_2F_1( { 3 \over2 } + \ell , 3 + \ell,  { 5 \over 2 }  ;- e^{- 2 \tau } ) \right] ~~~~~~~~~~
 \eea 
  Note that the imaginary terms come from the second term and start at order $e^{ - 3 \tau}$ which should be compared with the $\eta^3$ term in \nref{Soln}. In fact,  
  taking $\ell \to \infty $ and $\tau \to \infty $ with 
    ${ \ell \over \cosh \tau}  = - k \eta $ fixed, we recover the solution with the flat slices   \nref{flaWF}.     
   
     Of course, even with interacting fields we can get the wavefunction by evolving the full quantum field theory starting from $\tau = i \pi/2 $ and then going to real $\tau$. 
 However, the point we want to emphasize is that in the tree level approximation we can get the wavefunction by evaluating the action on a classical solution with the appropriate regularity conditions in  complex time. 
 
 In appendix \ref{Gend}, we write the solutions for general mass and dimension. 
  
  We can also write \nref{SolLt} as \cite{Ivo:2024ill}
  \begin{equation}
f_{\ell}^{\pm}(\tau)=\bigg(\frac{\ell+\cosh^{2}\tau + i \ell \sinh \tau}{\cosh^{2}\tau}\bigg)e^{- 2 i \ell \arctan e^{-\tau}}\,,\label{eq: f_l dS}
\end{equation}
 
 \section{The no boundary geometry for slow roll inflation } 
 \la{HHSlow}
 
 Hopefully, after the results of the previous section, the no boundary proposal \nref{NBp}
 seems natural. 
 %
 In pure de-Sitter the no boundary geometry is simply \nref{dSGlob} with $\tau$ starting at $i \pi/2$ and running to a large real value of $\tau$ along the contour in figure \ref{Contours}c.   
 The gravity action evaluated on this geometry is 
 \bea \la{ActGra}
 i I &=& i \left[ \int { R \over 2 }- V_c  - \int_{bdy} K \right] ~,~~~~~~~~~~{\rm with} ~~~~~~8 \pi G_N   =1
 \\ 
 & = & - 2 i \omega_3 { 3 \over V_c} \int_{i \pi/2}^{\tau_r} d\tau 3 \cosh \tau \sinh^2 \tau  = { 4 \pi^2 } { 3 \over V_c } \left[ 1 - i ( \cosh \tau_r)^3 + i { 3 \over 2}  \cosh\tau_r   + o(e^{ -\tau_r}) \right]  \la{PuredS}
  \eea
  where $\omega_3 = 2 \pi^2$ is the volume of $S^3$, 
   $V_c$ is the cosmological constant,  and we used that solution to the equations of \nref{ActGra} is de-Sitter with radius $R^2 = H^{-2} = { 3 \over V_c}$. Note that  $\tau_r \gg 1$ is a large value where we are setting the size of the scale factor.  This leads to 
  \be 
  |\Psi|^2 \sim  \exp ( 2 {\rm Re} [ i I ] ) = \exp\left( 8 \pi^2  { 3 \over V_c}\right) 
 \ee 
 In pure de-Sitter we can view this as just a normalization constant, which is useful only when we compare against other no-boundary geometries. This has been found to be the dominant geometry both for the bulk and the boundary. For example, for a boundary geometry with the shape of $S^1 \times S^2$ we can use an analytic continuation of the Schwarzschild dS solution as a no boundary geometry. The resulting action is smaller than for the three sphere \cite{Bousso:1998na,Conti:2014uda}.  
 
 We now discuss the no boundary geometry in the context of slow roll inflation. 
 This was discussed in \cite{Janssen:2020pii}, and our discussion is a small  improvement because we give analytic approximations. Numerical solutions were given in a number of papers, see e.g.  \cite{Hartle:2008ng}. 
  
 We consider gravity plus a scalar field with the action 
 \be \la{SrolI}
 I = \int { R \over 2}  - \half (\nabla \phi)^2 - V(\phi)  - \int_{\rm bdy} K ~,~~~~~~~~ 8 \pi G_N =1
 \ee 
 We then look for a spherically symmetric geometry and scalar field of the form 
 \be \la{Ansa}
 ds^2 = -dt^2 + a(t)^2 d\Omega_3^2 ~,~~~~~~~~~~~ \phi = \phi(t)
 \ee 
 They obey the equations 
 \bea \la{FriE}
  3 H^2 &=& 3 { \dot a^2 \over a^2 } = - { 3 \over a^2 } + \half \dot \phi^2 + V     \\ \la{SFeq} 
  0&=& \ddot \phi + 3 H \dot \phi + V'   ~,  
  \eea
  It is useful also to give the form of the equations in the slow roll approximation and for flat slices 
  \bea 
 ~~~~ 3 H^2 \sim V ~,~~~~~~~~~~~~~~~~~~~~~~~~~~3 H \dot \phi \sim - V' \la{SlowRoll}
  \eea
 
 We are interested in a solution where at $\phi_r$ we have a  3-sphere of a large size  $a_r$. 
 We will approximate the solution as follows. 
 First we neglect the $1/a^2$ term in the Friedman equation \nref{FriE}. After we do this,  the equations depend only on $a/a_r$.  We can further approximate the equations using the standard slow roll approximation \nref{SlowRoll}. If we follow this standard slow roll solution, we  find that at some point the size of the sphere ``crosses the horizon''. What this means is, by definition, that 
 \be 
   H_* \equiv H(\phi_*)=  { 1 \over a_*}  = { 1 \over a_r } e^{ {\cal N}(\phi, \phi_* ) } \la{HHhor} 
  \ee 
  we denoted by $\phi_*$, $H_*$ and $a_*$  the   values of $\phi$, $H(\phi)$ and $a$  where this happens. In the last expression we emphasized the fact that $\phi_*$ depends on $a_r$ and $\phi_r$ and we expressed $a_*$ in terms of the number of e-folds accumulated between between the values $\phi_*$ and $\phi_r$ for the inflaton. 
   
    Around this time, \nref{HHhor}, the zeroth order approximation to the no boundary solution is simply de-Sitter with radius $H_*^{-1} $  
  \be \la{dSgsL}
  ds^2 = { 1 \over H_*^2 }  ( - d\tau ^2 + \cosh^2 \tau d\Omega_3^2 ) ~,~~~~~~~{\rm with } ~~~~H_*^2 \sim V_*/3 ~,~~~~~~~ \tau = H t 
  \ee 
  where we also related $\tau$ to the proper time $t$ in \nref{Ansa}. 
 In this approximation,  the complex time contour for the no boundary solution starts again from $\tau = i \pi/2$.  
  
 We can improve this approximation by solving the equation for the scalar field in this geometry. We write the scalar field as $\phi = \phi_* + \varphi$ and we expand the equation for the scalar field  \nref{SFeq}  as
 \be \la{varEq}
 0= \partial_\tau^2 \varphi  + 3 \tanh\tau \partial_\tau \varphi + { V'_* \over H_*^2 }  
 \ee 
 Note that we expanded the potential to zeroth order in $\varphi$ and not to first order, since the zeroth order already   gives us the dominant contribution. 
 The general solution of \nref{varEq} that is regular at $\tau = i \pi/2$ is 
 \be \la{varphso}
 \varphi = { V'_* \over V_* } \tilde \varphi = { V'_* \over V_* } \left[ { 1 + i \sinh \tau \over \cosh^2 \tau}     -   \log[ 1-i\sinh \tau   ]  + c \right] ~,~~~~~~~~c = -   i { \pi  \over 2} 
 \ee 
 where $c$ is a constant that was determined as follows.  
 We  expand $\varphi$ for large $\tau $ to find 
 \be 
 \varphi = { V'_* \over V_* } \left[ - \tau  + \log 2 + i \pi/2 + c +\cdots  \right] 
 \ee 
 We expect that we should match this to the slow roll solution, expanded around $\phi_*$, which is   
 \be 
 \phi = \phi_*  - { V_*' \over V_* } \log (a/a_*)   
 \ee 
  where $\log a/a_* = H_* t $, with $t$ proper time in the slow roll solution with flat slices (which is not the same as the proper time in \nref{dSgsL}). 
  Since $a = { 1 \over H_*} \cosh \tau \sim { 1 \over H_* } {e^t \over 2} $, 
    we should identify $\log(a/a_* ) \sim  \tau - \log 2 $.  This then gives  $c= - i \pi/2$. The solution  \nref{varphso} matches the series expansions obtained in  \cite{Janssen:2020pii}.

  It is also useful to note that the profile of $\varphi$ at large $\tau $ becomes mostly real 
  \be 
  \varphi \sim { V'_* \over V_* } \left[  - \tau + \log 2  + 3 e^{ - 2 \tau } - i { 16 \over 3} e^{ - 3 \tau } + o (e^{-4 \tau } )\right] \la{vfLa}
  \ee 
  as $\tau \to \infty$, but  there is a small decaying imaginary part.   
  
  In principle, we can also compute the correction for the scale factor $\delta a$, see appendix \ref{ScaleFactor}. However, we will describe the deformation of the metric in a slightly more direct way by taking $a$ to be the time variable. 
  In other words, we write the metric as 
  \be \la{MeA}
  ds^2 = g_{aa} da^2 + a^2 d\Omega_3^2 
  \ee 
  Writing the action \nref{SrolI} in these variables we find that $g_{aa}$ appears only algebraically. We can find its equation of motion
  \be 
 g_{aa} =    {   3   -   \half \left( a { d\phi \over d a } \right)^2    \over {     3  - V(\phi) a^2 } } \la{metra}
 \ee
   and substitute back into the action to find 
  \be \la{Laga}
  I = - 2 \omega_3 \int_0^{a_r} da a^2 \sqrt{ 3   - \half   \left( a{ d\phi \over d a } \right)^2  } \sqrt{  V(\phi) -{  3 \over a^2}  }  
 \ee  
  We can now  rewrite \nref{varphso} in terms of $a$ as  
 \be \la{varphsoa}
 \varphi = { V'_* \over V_* } \left[  { 1 + i \sqrt{ \alpha^2 -1} \over \alpha^2 } - \log \left[ 1- i  \sqrt{\alpha^2 -1} \right]  - i { \pi \over 2}  \right] ~,~~~~~~~ \alpha \equiv  a H_* = \cosh \tau 
 \ee 
We can now get the metric to order $\epsilon  = { V_*'^2 \over 2 V_*^2 } $ by expanding \nref{metra}. Note that we expand the numerator to second order in $\varphi$ and the denominator only to first order, because the derivative of the potential is already small. We find  
 \bea 
g_{aa} &=&  
{ 1 \over 1 - \alpha^2  }\left\{ 1  + { V_*'^2 \over V_*^2 } \left[ - { 1 \over 6 } { ( \alpha  d_\alpha \tilde \varphi)^2  } + {  \alpha^2  \tilde \varphi \over ( 1- \alpha^2 ) } \right] \right\}~,~~~~~~\varphi = { V'_* \over V_*} \tilde \varphi 
\la{fulco}\\ 
&=& { 1 \over 1 - \alpha^2  }\left\{ 1  + { V_*'^2 \over V_*^2 } \left[  \log \alpha   - {1 \over 6 } + {( \log  \alpha   - 5/3) \over \alpha^2 } + i { 4 \over 3 } { 1 \over \alpha^3 }  + o (1/\alpha^4)  \right] \right\} \la{ResIm}
\eea 
 Where we   inserted the solution \nref{varphso} in the form \nref{varphsoa}. In the second line, we have expanded the correction for large $\alpha$, which is late times. We see that there is also a small imaginary correction. 
 
 Note that the leading order expression for the metric 
 \be \la{leabmP}
 g_{aa} = { 1 \over  1 -\alpha^2} 
 \ee 
 has a pole at $\alpha=1$ where the signature changes. We should go around this pole in the upper half plane. This is simply the point $\tau =0$ in the variables \nref{dSGlob} and setting 
 $\alpha = \cosh \tau $ we obtain \nref{dSGlob} from \nref{leabmP} \nref{MeA}.

The expression \nref{fulco} contains a double pole at $\alpha =1$ whose coefficient has a negative imaginary part. This suggests that the exact answer is such that the pole at $\alpha =1$ of \nref{leabmP} is shifted by an order $\epsilon$ amount towards the upper half plane, though the order at which we worked is   not enough to prove this. 
 However, this explains why if one attempts to solve the equation numerically by restricting to real values of $a$ one is not able to find the right solution. In numerical solutions, we start from the equations of motion of \nref{Laga} and solve them along a complex contour in the upper half  complex $a$ plane that stays away from the $a=\sqrt{3/V}$ (or $\alpha =1$) region. We look for a  solution with $\partial_a \phi =0$ at $a=0$ and some value $\phi(0) $ which we need to tune in order to obtain a solution with $\phi(a_r)  =\phi_r$, where $a_r$ and $\phi_r$ are the given values at the boundary.    
 
 We should emphasize that the solutions obtained above are a good description around the region where we make the excursion into complex metrics and the sphere shrinks. 
 As we go to large $\tau$ they then start matching   the usual slow roll solutions. More precisely,   we should match the large $\tau$ solutions in \nref{ResIm} and \nref{vfLa} to the   slow roll solutions with flat slices plus some small deformations due to the spatial curvature, see appendix \ref{HJTheory} for more details.

   We can now evaluate the on shell action. The first observation is that the extrinsic curvature term in \nref{ActGra}  cancels out after we integrate by parts and take the action to \nref{Laga}. So we only need to evaluate \nref{Laga}. 
   
   We will first evaluate the action in the region around the value $\phi \sim \phi_*$. In that case, as a first approximation we can simply say that $\phi = \phi_*$ and then consider the correction $\phi = \phi_* + \varphi$ and expand in $\varphi$. We expand the kinetic term in $\phi$ to second order in $\varphi$ and the potential to first order. These two terms are of the same order in the slow roll factor $\epsilon$. 
   
   After we do this, we find the action 
   \bea 
   i I &=& - 2 i \omega_3 \int_0^{a_{r} } da a  3 \sqrt{  { V_* \over 3 } a^2 -1  } \left[ 1 - { 1 \over 12 } (a \partial_a \varphi)^2 + \half { V'_* 
   \over V_* } { \varphi  \over 1 - { 3 \over V_* a^2 } }  \right] 
   \\
   &=& -   i \omega_3 { 3 \over V_* }  \int_0^{\alpha_{r} } d (\alpha^2)  3 \sqrt{ \alpha^2 -1  } \left\{ 1 + \epsilon \left[ - { 1 \over 6 } (\alpha \partial_\alpha \tilde \varphi)^2 +   { \alpha^2 \tilde \varphi  \over \alpha^2 -1 }  \right] \right\} \la{SecExpA}
   \eea 
  where we defined $\alpha = H_*  a = \sqrt{ V_*/3} a$. 
  
  The leading order term, obtained by setting $ \epsilon =0$ in  \nref{SecExpA}, just gives 
  \be \la{ActC}
  i I = - i \omega_3 { 3 \over V_* } 2 \left. (\alpha^2 -1)^{3/2}\right|_{0}^{\alpha_r} =   { 4 \pi^2} { 3 \over V_* } \left\{ 1 - i \left( \alpha_r^3 - { 3 \over 2 } \alpha_r + o(1/\alpha)    \right)   \right\} 
  \ee 
 Of course, this is just the purely de-Sitter result \nref{PuredS}. The real part of $iI$  comes from analytically continuing the square root along the upper half $\alpha$ complex plane to $\alpha =0$. The purely imaginary terms can be viewed as  local terms along the late time slice which involve the integral over the volume or the volume times the scalar curvature. These imaginary terms are important to get the right extrinsec curvature of the classical solution. The extrinsec curvature is related to  a derivative with respect to $a$ of the wavefunction $K = 3 H \propto  i {1 \over a^2} { d \over d a  } \log \Psi = - { 1\over a^2 } {d I \over d a} $.  In \nref{ActC} we evaluated the action assuming $\alpha_r$ is not too large, so that we can use the approximation where we perturb around a constant $\phi_*$.

 The action of the full inflationary problem includes the correction of order $\epsilon$ in \nref{SecExpA} which is small and not too relevant for the discussion, but  we give the answer in appendix \ref{Epsilonterms}. It is worth noting that this correction does not lead to any real term in $iI$ that grows with $\alpha_r$.   This is important because we could have worried that the imaginary terms going like $i /a^3$ in $\phi$ \nref{vfLa} or the metric 
 \nref{ResIm}  could have led to a correction, which when multiplied by the leading term going like $a^3$, could give an extra contribution that is real to $i I$.  
   
    In the full no boundary geometry, we are interested in making $a_r$ exponentially large, which means that we have a large number of e-folds. This takes us away from the regime where the action \nref{SecExpA} was evaluated, where $\alpha$ was large but not exponentially large. 
    
    In order to find the action in the regime of interest,  it is useful to use the Hamilton Jacobi equation which is an equation for the action\footnote{ Of course, this Hamilton Jacobi equation is the semiclassical approximation of the Wheeler de-Witt equation.}. We can solve this equation for the exponentially large terms in $a$ and find that the action takes the form, see appendix \ref{HJTheory}, 
 \be \la{AsCt}
 i I   = i { 4 \pi^2 } \left\{-  [H_f(\phi_r)   a_r^3 - { 3 \over 2 } { a_r B(\phi_r) } ]   + o(\log a_r)  \right\}
 ~,~~~~~~~
 \ee 
 where $H_f(\phi)$ is the value of the Hubble constant as a function of $\phi$ for the solution with no curvature, with flat slices. $B(\phi)$ is a function fixed by an equation that we give in \nref{EqBb},  whose solution goes like 
 \be 
 B(\phi_r) = { 1 \over  H_f(\phi_r) } + \cdots 
 \ee 
 to leading order in the slow roll factors.

 We can further argue that the subleading term should take the form 
  \be \la{ActFi}
 i I = { 4 \pi^2 } \left\{-  i [H_f(\phi_r)   a_r^3 - { 3 \over 2 } { a_r B(\phi_r) }]  +  { 3 \over V_* }   \right\}
 \ee  
 This is argued as follows. We know that $H$ is related to the momentum conjugate to $a$ via 
 \be 
4 \pi^2   H =  - { 1 \over 3 a^2 } \partial_a I  
\ee 
We then compare the expression for $H$ that we obtain from \nref{ActFi} with the one that we get by solving the equations approximately and see that they match.  We give the details in appendix \ref{HJTheory}.  
The claim is that \nref{ActFi} is correct up to terms of order $\epsilon$.

 

  As a final comment, note that  the  imaginary part of the correction for the metric    in \nref{ResIm} leads to a possible violation of the   Kontsevich-Segal-Witten (KSW)  criterion for an allowable complex metric \cite{Louko:1995jw,Kontsevich:2021dmb,Witten:2021nzp}. This was noted in numerical solutions in \cite{Hertog:2023vot}. In appendix \ref{KSW},  we give a simple explanation and show that it leads to a constraint bounding $\epsilon$ in terms of the number of e-folds 
   \be \la{Boundr}
    r = 16 \epsilon \leq { 8. \over {\cal N } } 
    \ee 
    which summarizes the findings of a more detailed numerical analysis in \cite{Hertog:2023vot}. 
    It is not clear what we should make of this, given that there are some situations where metrics violating the KSW criterion are giving physically reasonable answers \cite{Chen:2023hra}\footnote{We should probably view the KSW criterion as good evidence that a metric is allowed.  But if the criterion is not obeyed, then  one needs more work to argue whether the metric is really  allowed or not.}. Note that given the big problem with the Hartle-Hawking proposal reviewed around \nref{Shife},  it seems premature to try to apply \nref{Boundr} to phenomenology\footnote{Though it is a bound in agreement with observations.}.

 \section{Comments on the connection to AdS/CFT} 
 
 In the context of AdS/CFT we encounter a similar geometry. In particular, if we consider a euclidean three dimensional field theory on a  three dimensional geometry specified by the metric $h_{ij}$, then we should consider a euclidean four dimensional geometry with a boundary given by this three dimensional geometry. Then the classical action is  an approximation to the partition function of the field theory \cite{Witten:1998qj,Gubser:1998bc}\footnote{Each of the two sides should be viewed as defined up to the addition of local counterterms \cite{Henningson:1998gx}.}
 \be \la{AdSCFT}
  Z[h_{ij}] \sim  e^{ - I_E}   
  \ee  
  In this case,  the standard gravity computation gives results that are in agreement with the field theory answer, when the latter can be computed.   For a CFT$_3$,  and   when the three manifold is a 3-sphere,  the quantity in the left hand side of \nref{AdSCFT} is defined to be the ${\cal F}$ function of the field theory \cite{Kapustin:2009kz,Jafferis:2010un,Jafferis:2011zi,Casini:2012ei,Drukker:2010nc}
  \be 
  {\cal F }  = - \log Z  ~,~~~~~~~~~ \log Z \sim  - I_E = - {  2\pi R_{AdS}^2 \over 4 G_N} = -{ 4 \pi^2 } { 3 \over |V| } 
 \ee 
 which we see has the same form as the leading term of \nref{ActC}, except for the overall sign. In fact, it the same as what we analytically continued $V$ in \nref{ActC} to negative values.  
  
 It is amusing that in \cite{Hartle:1983ai} it was believed that the no boundary proposal applied to $dS$ but not to $AdS$, while now we think that that it applies beautifully for euclidean $AdS$ but we are puzzled about the $dS$ case. 
  
  Of course, a possible dS/CFT correspondence \cite{Witten:2001kn,Strominger:2001pn}
   would give us  a way to compute the wavefunction of the universe from a quantum field theory\footnote{Note that objections to dS/CFT based on the thermal fluctuations \cite{Dyson:2002nt}, or the decay of de Sitter, are circumvented by viewing dS/CFT in the context of computing the wavefunction \cite{Maldacena:2002vr}, which computes the probability that the universe has not yet decayed, and where the boundary geometry takes a specific value.}. Given a specific candidate for a CFT dual to de Sitter, we can of course compute the sphere partition function an compare to the gravity answer. 
   For an example involving an exotic theory of gravity,  see \cite{Anninos:2011ui,Anninos:2012ft}.    
  
  \section{Proposals for dealing with the problem } 
  
  \la{Solutions}
  
  There are various proposals for dealing with this problem as it concerns our universe. In this section, we attempt to summarize these discussions,  but readers should look at the references to form their own opinion.

  \subsection{Slow roll eternal inflation and the Hartle Hawking proposal}
  \la{Starob}
   
 We   start by describing a situation where we can obtain the Hartle Hawking answer using a different method.   
  This is the problem of slow roll eternal inflation,  where the inflationary potential has a very small slope,  so that the classical motion over an e-fold is comparable to the quantum fluctuations. This happens when $ \epsilon { M_p^2 \over H^2 } \sim 1$, which is   when the amplitude of the curvature fluctuations become of order one,  as computed via \nref{Sfc}. 
 
  Starobisnki has proposed analyzing this problem as follows \cite{Starobinsky:1986fx}. Imagine that you are an observer moving along a timelike geodesic in this spacetime. The idea is that the quantum fluctuations would appear as   random fluctuations in the scalar field. Then  the evolution of the probability distribution for the scalar field obeys the Fokker Planck equation. 
   Namely, we start with the equation for the scalar field with a random source  \cite{Starobinsky:1986fx}
  \be \la{Stev}
  \dot \phi = - { V' \over 3 H } + f(t) ~,~~~~~~~~~~
  {\rm with }~~~~~~~ \langle f(t) f(t') \rangle =  { H^3 \over 4 \pi^2 }   \delta(t-t') 
  \ee 
  This source is chosen so that, on its own,  it gives rise to $ \langle \phi (t) \phi(0) \rangle \sim { H^2  \over 4 \pi^2 }  \times H t  $, so as to match the form of the quantum expectation values of the fluctuations in position space $\langle \phi(x) \phi(0) \rangle \sim -{H^2 \over 4 \pi^2 } \log x $, after we identify $\log x \to - Ht $,  which is the number of e-folds. 
  
  The stochastic evolution \nref{Stev} leads to 
 the Fokker Planck equation for the probability $P(t,\phi)$  that the field has a value $\phi$   \cite{Starobinsky:1986fx,Goncharov:1987ir}

  \be 
  \partial_t P(t,\phi) + \partial_\phi J_\phi(t,\phi) =0 ~,~~~~~~~~~~~
  J_\phi(t,\phi)  = - {  V' \over 3 H } P(t,\phi)  - \partial_\phi \left( { H^3 \over 8 \pi^2 }  P(t,\phi)    \right)   
  \ee 
  where $J_\phi$ is the probability current. What is interesting for us is that if the potential is bounded below $V(\phi) \geq V_{min} > 0 $ then we get an equilibrium configuration with $J_\phi =0$, which, using $H^2 \sim V/3 \ll 1$   is 
  \be \la{ProbSta}
  P \propto \exp\left(  8
  \pi^2  { 3 \over V(\phi) }\right)
    \ee 
    which  is indeed equal to the Hartle Hawking distribution \nref{HHInt}. 
    
    More generally, studies of bubble nucleation and external inflation also lead to rates which are consistent with  \nref{ProbSta}, as long at the potential has a positive minimum \cite{Lee:1987qc}.  
    
    We can think of this as a thermal distribution centered on the minimum of the potential and giving us the probabilities of the other vacua as thermal fluctuations away from the minimum cosmological constant de-Sitter space \cite{Lee:1987qc}.   
        
    So, this is indeed a context where the Hartle-Hawking answer is indeed reasonable. 
  
  

   \subsection{Perhaps the path integral is ill defined because gravity is not renormalizable or the action is not bounded below}
   
   The gravitational path integral has two problems. One is that the theory is not renormalizable, the other is that the sign of the Euclidean action is not bounded below, which poses a problem for the Euclidean theory and is ultimately responsible for the sign in \nref{HHInt}. 
   
   Semiclassical gravity makes sense as an effective field theory and we expect that, at least in some special cases,  it can be UV completed to a string theory.  The fact that that action is not bounded below is dealt with using a suitable choice for the integration contour \cite{Gibbons:1978ac}. 

   In many cases,  Euclidean gravity leads to saddles that give physically reasonable answers such as the Euclidean black holes or euclidean AdS spaces.   More recently, the euclidean path integral has been shown to encode subtle aspects of the fine grained entropy of evaporating black holes, see \cite{Almheiri:2020cfm} for a review. 
   
   \subsection{Perhaps we are not in the Hartle Hawking state} 
   
   \la{NoHHState}
   
   The reader is probably puzzled by the following. It is usually said that slow roll inflation implies that the universe is spatially flat, and that this is a generic prediction of inflation, which is in very good agreement with observations. On the other hand, we said here that the no boundary geometry  predicts a positively curved universe in seemingly the same context of slow roll inflation. 
   
   The point is that the usual inflationary discussion {\it assumes} that we started somewhere high along the potential in the past, with a suitable initial conditions \cite{East:2015ggf}. Then the universe expands leading to the standard inflationary solution. 
   
   On the other hand, the no boundary proposal is more ambitious, it is a theory of the initial conditions. More precisely, it rephrases the problem in terms of a prescription for computing the wavefunction for observables we can observe now. It is supposed to compute all the ways in which the universe can end up looking like it is now. Note that we are viewing the Hartle Hawking proposal not as a probability that the universe would start somewhere, but as the probability that the universe will end up looking like it looks at the final surface where we evaluate the wavefunction. 
   
   However, it could be that the Hartle Hawking state is just one possible wavefunction in quantum gravity and that there are others. And that for some reason we are not in the Hartle Hawking one. One reason for not being in the Hartle Hawking one could be that the potential has regions where it is zero or negative, which   makes the Hartle Hawking state not normalizable. This is what is expected in the string landscape \cite{Kachru:2003aw}, for example. 
  
  Note that simply saying that the inflaton starts at some point high in the potential, is the same as saying that this is   not the Hartle Hawking state, since we are projecting onto special  solutions where  the scalar field indeed went through that high point in the potential.

  The Starobinsky discussion of section \ref{Starob} suggests that the Hartle-Hawking state is   a particular state in thermal equilibrium. It seems that our  universe is in a different state,  which is out of equilibrium.  
   This is in fact, the resolution of a similar paradox: the Olbers paradox. Particular models in this direction were discussed in \cite{Bousso:2011aa,Harlow:2011az} and  consider a more realistic landscape with terminal decays (decays into anti-de-Sitter space, for example).

   We should also mention that a picture of eternal inflation with the universe eventually tunneling to a region with usual slow roll inflation would more naturally create a situation with negative curvature \cite{Coleman:1980aw,Bucher:1995ga}, as opposed to a universe with positive curvature .

   \subsection{Perhaps we should consider a tunneling wavefunction  } 
   
  One could imagine flipping the sign of the action, or its imaginary part, by moving in the opposite direction in the complex plane, see e.g. \cite{Linde:1998gs}.
  This also has the consequence of changing the sign for the wavefunction of the small fluctuations, which was correct with the Hartle Hawking proposal but would now be wrong  \cite{Hawking:1998xk}.
  
  A closely related idea is that the same geometry has a tunneling interpretation \cite{Vilenkin:1982de},   where one imagines that the universe starts very small and tunnels to a size of order $H^{-1}_*$. If one considers then the probability that we tunnel to a universe which is a sphere plus fluctuations, then again we would get the wrong sign for the fluctuations, if we use a small deformation of the geometry we used to compute the tunneling to  a round sphere.        In \cite{Vilenkin:2018dch} some proposals were made to avoid this result. These proposals amount to imposing boundary conditions at $a=0$ which are not the no boundary conditions and treat this point in the geometry as a special point. While this might ultimately be correct, it is not the no boundary geometry and it might be related to the points in section \ref{NoHHState}, where we start from a different state. 
   
   It was argued  in \cite{Feldbrugge:2017kzv,Feldbrugge:2017mbc} that the Vilenkin sign is correct     if one starts from a Lorentzian path integral.  The opposite sign was argued for in \cite{DiazDorronsoro:2017hti} and criticized in \cite{Feldbrugge:2018gin}.

    \subsection{Perhaps we should apply a selection principle} 
    
   Perhaps we are very rare as observers so that there is a very small probability $p$ per unit volume that we exist. 
    In that case, it seems advantageous to make a bigger universe so that we get an expression of the form \cite{Hartle:2010vi,Hartle:2010dq}
    \be \la{FpV}
    p {\cal V} \exp\left(  { 8 \pi^2  } { 3 \over V_*}\right) 
    \ee 
    for the probability that there is at least one observer in the total volume, ${\cal V}$, of the universe  (assuming $p {\cal V} \ll 1$). 
    Notice that ${\cal V} \propto \exp( 3 {\cal N } )$  where  ${\cal N}$ is the number of e-folds. This volume factor increases when we increase the number of e-folds, while the second factor decreases, see \nref{Shife}. %
    Extremizing \nref{FpV} over ${\cal N}$ would give us a $V_*$ in the regime of eternal inflation\footnote{The precise condition gives us ${\dot \phi^2 \over H_*^4 } = 3/(8 \pi^2)$. This is on the eternal inflation side of the boundary found in    \cite{Creminelli:2008es}, which is ${\dot \phi^2 \over H_*^4 } = 3/(2 \pi^2)$.}. This means that for \nref{FpV} to dominate over the solution with $\phi_*$ corresponding to  60 efolds, we would need to have 
    \be \la{pUnr}
     p < \exp\left( - { 8 \pi^2  } { 3 \over V_{60}}\right) \sim \exp\left( -{ A_s \over 2 \epsilon } \right) \sim \exp( -10^{11} ) 
     \ee 
     Of course, here we are not taking into account the possibility of setting the scalar field at the present minimum, which would give the exponential of the cosmological constant problem. 
   But even ignoring that issue, we find that $p$ is fairly small. So, what selection effect could give rise to it? 
   
   First note that all inflationary trajectories we consider end up in the standard model, so we are not changing the laws of physics, we are only changing the overall curvature of the universe. 
   If the probability for the emergence of observers somewhat like us is smaller than \nref{pUnr}, then this idea could be reasonable. 
   
   However \cite{Hartle:2010vi,Hartle:2010dq} proposed that any observation of more than around  $10^{11}$ bits would already place us as unlikely observers. In other words, \cite{Hartle:2010vi,Hartle:2010dq} propose that    
      by simply filling a computer hard drive with random numbers you already become an unlikely observer. We do not find that this is a reasonable way to apply a selection principle, since the contents of that computer drive  would be a typical result among all the results you could have obtained regardless of the volume of the universe. 
     
     By the way, notice that these selection effects are saying that the philosophy behind figure \ref{Reheating} is wrong, since the presence of special type of observers in the later universe modify the distribution.    

\subsection{Perhaps the quantum corrections become important } 

One could wonder whether the one loop quantum corrections could become important enough to balance the classical effects. This was important for addressing the entropy of evaporating black holes \cite{Penington:2019npb,Almheiri:2019psf}. In our case, the quantum corrections seem to be small around the time when the universe was complex. As it approaches the Lorentzian solution one expects that they should be constrained by unitarity. In other words, for a real lorentzian universe the probability should be independent of where we evaluate it along the inflationary trajectory. In the first approximation,  where we approximate the no boundary geometry by de-Sitter space, the quantum correction are related to computing the norm of the usual de-Sitter vacuum for the quantum fields, which is constant and independent of the time at which we evaluate it. So it seems unlikely that quantum corrections could change the answer\footnote{There are some special no boundary geometries were quantum corrections indeed   balance   classical effects, see 
\cite{Maldacena:2019cbz} for an example involving an $S^1 \times S^2$ boundary with a very large $S^1$.  }. Nevertheless, it would be interesting to investigate this question further. 

 \section{Final Comments} 
 
 The no boundary proposal is a very natural and mathematically elegant proposal for the wavefunction of the universe and it is unsettling that it disagrees with observations. 
 
 It is possible that we are not applying it properly.  Or that we should use another   principle that selects the right state for the universe. This seems to be the majority view among quantum cosmologists.

 \subsection*{Acknowledgments}

We would like to thank O. Aharony, T. Hertog, T. Jacobson and E. Witten for discussions. 

J.M. is supported in part by U.S. Department of Energy grant DE-SC0009988.

\appendix

\section{Solutions in global coordinates that are relevant for evaluating the wavefunction } 
\la{Gend}

In this appendix, we write the solution of the wave equation for a massive field with mass $m$ in general $dS_d$ that is relevant for evaluating the wavefunction. This is the solution regular at $t=i \pi/2$. Setting $ R_{dS}=1$, and defining $\mu = \sqrt{ m^2 -(d-1)^2/4} $,  we get 
\be \la{SolReg} 
\phi  \propto (\cosh t)^\ell   e^{  ( \ell + { d -1 \over 2 } - i \mu ) t }  ~_2F_1 \left( { d -1 \over 2 } + \ell , { d-1 \over 2 } + \ell - i \mu , d-1 + 2 \ell ; 1 + e^{  2 t } \right) 
\ee 
 In this form, it is manifest that it is regular near $\tau =i\pi/2$, we can go real $\tau > 0$ by using the usual rules for the transformation of hypergeometric functions. 
 Setting $d=4$ and $ \mu = i (d-1)/2$ we recover \nref{HypSph}.

\section{Correction to the scale factor }
 \la{ScaleFactor}
 
 Working in the metric \nref{Ansa} we can expand the solution around the region where $\phi \sim \phi_*$. 
 We expand  the scale factor of the metric \nref{Ansa} around the leading metric \nref{dSgsL} as   
 %
 \be 
 a = H_*^{-1} \cosh \tau ( 1 + \gamma +\cdots  ) ~,~~~~~~\gamma \sim o(\epsilon) 
 \ee 
 We first set $\gamma =0$, we  
  expand $\phi$ around $\phi_*$ as $\phi = \phi_* + \varphi$,  and obtain the solution \nref{varphso}. 
 We then expand the Friedman equation \nref{FriE} in $\gamma$ to obtain the equation 
 \bea 
  \left[ \tanh \tau \partial_\tau \gamma - { 1 \over \cosh^2 \tau } \gamma \right] &=&  \tanh^2 \tau { d \over d\tau } \left[ {\gamma \over \tanh \tau } \right]= { V_*'^2  \over 2 V_*^2  } 
h(\tau)   \la{LorAeq}
\\ 
h(\tau) &\equiv & { 1 \over 6 } \tilde \varphi'(\tau)^2 + \tilde \varphi(\tau)  
\eea
 We can also write the equation in $\theta $,  with $t = i \theta$. In that case we find that $\varphi( i\theta)$ , and $h(i \theta)$ is regular at $\theta =   \pi/2$ thanks to the boundary conditions we have chosen. 
  We integrate   equation \nref{LorAeq} as 
  \be \la{gamEx}
  \gamma(\tau) =   { V_*'^2  \over 2 V_*^2  } \left[ \tanh \tau \int_{i \pi/2}^\tau d\tau' { h(\tau') \over \tanh^2  \tau' }  + c \tanh \tau \right] ~,
  \ee 
  where we integrate along a contour in the upper half plane. Notice that the first term is regular near $\tau =  \pi/2$. One would be tempted to drop the second term on the grounds that it is singular at $\tau=0$. However, it is more convenient for us to keep it. 
  Doing the integral explicitly we get 
  \bea
\delta \log a &=& { {V}_*'^2 \over 2 V_*^2} \left[- 2 \tanh t {\rm Li}_2( - i e^{- \tau} ) + 2 \log[ 1 + i e^{-\tau}] - { 2 i \over 9 } {(1 + i \sinh \tau )^2 \sinh \tau \over \cosh^4 \tau } - { 20 \over 9 } i { \sinh \tau \over \cosh^2 \tau }+ \right.
\cr 
 & ~& \left. + { 20 \over 9 } \tanh^2 \tau + (\log 2 - { 5 \over 6 })  \tau \tanh \tau + ( - {\tau^2 \over 2} + i { 5 \over 12} {\pi }   - {7 \pi^2 \over 24 } - i {\pi \over 2 } \log 2) \tanh \tau  + \right. 
 \cr 
 & ~& \left. + \tau - {1 \over 3} - \log 2+ c \tanh \tau \right]
 \eea 
 We can now expand this around large $\tau$ to obtain 
 \be \la{LargeTime}
\gamma  = { V_*'^2 \over 2 V_*^2 } \left\{  - {\tau^2 \over 2} + ( {1 \over 6 } + \log 2) \tau + { 17 \over 9 } + i { 5 \pi \over 12} - { 7 \pi^2 \over 24} - \log 2 ( 1 + i { \pi \over 2 } )   + c + o(e^{ - 2 \tau } ) ] \right\}
\ee 
  Let us first understand the origin  of the quadratic and linear terms in $\tau$. They can be viewed as arising as follows. 
 The quadratic term in $\tau$ in \nref{LargeTime} 
arises from the fact that the expansion rate, $H$, 
is changing with time.   Since $H \propto \sqrt{V}$, we have 
\be 
{ \partial_\tau H \over H } = { \dot H  \over H_*^2 } = -  \half { (V')^2 \over   V^2 }  = - \epsilon 
\ee 
This means that $H = H_* ( 1 - \epsilon \tau) $. Then we have 
\be 
da = H dt = { H \over H_* } d\tau = (1 -\epsilon \tau ) d\tau = d ( \tau - \epsilon { \tau^2 \over 2 }  ) 
\ee 
this agrees with \nref{LargeTime}, including the sign. 
The linear in $\tau$ term in \nref{LargeTime} comes from the change in $H$ due to the $\dot \phi$ term. 
\be 
3 H^2 = \half \dot \phi^2 + V \to  H = \sqrt{ V\over 3} \left( 1 + { V'^2 \over V^2 } + \cdots \right) 
\ee 
where the correction comes from the kinetic term and we used the slow roll form of the equations. This gives us the term linear in $\tau $ in \nref{LargeTime} with the right coefficient. 

We want to define things so that in the large $\tau$ region we match the standard inflationary solution with flat slices which is expected to be of the form 
\be \la{FlaSli}
\gamma_{flat} \sim  { V_*'^2 \over 2 V_*^2 } \left\{  - {(\Delta N)^2  \over 2} + {1 \over 6 } \Delta N     \right\} ~,~~~~~~~~\Delta N = \tau - \log 2 = \log \left( { a \over a_* } \right)
\ee 
where $\Delta N$ is the number of efolds in the flat slicing from the value of $a$ to the one at $a_*$ where \nref{HHhor} is obeyed. Now the idea is that there is no constant term in \nref{FlaSli}. So we can now determine the value of $c$ by matching \nref{FlaSli} with \nref{LargeTime} and demanding that there is no constant term. 
We find 
\be \la{cvald}
c = - \half (\log 2)^2 + { 5 \over 6 } \log 2 + i { \pi \over 2}\log 2  - { 17 \over 9 } - i { 5 \pi \over 12 } + { 7 \pi^2 \over 24 } 
\ee 
Now, we need to explain how to interpret the singularity at  $\tau = i  \pi/2$ from the term proportional to $c$ in \nref{gamEx}. We can remove the singularity by changing the origin of time.   In other words, the idea is that the scale factor is   non-singular at 
\be 
\tau = i { \pi \over 2}  - \epsilon c \la{Shiftct}
\ee 
since the last term in \nref{gamEx} is what we get by expanding 
\be 
 a = H_* \cosh (\tau + \epsilon c ) = H_* \cosh \tau \left[ 1+ \epsilon \tanh \tau   c \right] 
 \ee 
 Notice that the shift \nref{Shiftct} has both real and imaginary parts \nref{cvald}. It is also of order $\epsilon$ so that it does not affect the solution for $\varphi$ which was of order $\sqrt{\epsilon}$ \nref{varphso}. 
When we described the geometry using the scale factor as time, as in \nref{MeA} \nref{fulco}, 
  we did not have to worry about such shifts \cite{AMS}, and we did not need to do any integral   to obtain $g_{aa}$. Of course, we could also have obtained $\gamma$ by starting from \nref{fulco} and used the change of variables for the time coordinate from $t$ to $a$. 
 
  For similar reasons, we  also found it convenient to use the parametrization in \nref{MeA} when solving numerically the equations \nref{FriE} \nref{SFeq}  for the no boundary conditions.  

 It is also useful to expand $\gamma$ to higher orders as 
 \be 
 \gamma = { {V}_*'^2 \over 2 V_*^2} \left\{ - {(\tau- \log 2)^2 \over 2 } + { \tau - \log 2 \over 6 } +   [ (\tau -\log 2)^2   +  {5 \over 3 }  ( \tau -\log 2)  - { 11 \over 6 }  ]e^{- 2 \tau  } + i { 32 \over 9 } e^{ - 3 \tau } + \cdots  \right\} \la{Imava}
  \ee 
   We note that the first imaginary term arises at order $e^{ - 3 \tau }$, as we also had for $g_{aa}$ in \nref{ResIm}.

   \section{Evaluating the action using the Hamilton Jacobi equation} 
   \la{HJTheory}
   
   In this appendix, we want to give a few more details on the Hamilton Jacobi equation and the value on the on shell action. This was introduced in \cite{Salopek:1990jq} to study aspects of inflation and it is directly connected to the semiclassical limit of the minisuperspace Wheeler de Witt equation. 
   
   We start with the action \nref{Laga}. For convenience we drop the factor of $ \omega_3$ which we will restore at the end. So we work with $\tilde I = I/ \omega_3 $. We view $a$ as time and $\phi(a)$ as the dynamical variable. 
   
   We then can compute the momentum conjugate to $\phi$ and the Hamiltonian ${\cal H}(p,\phi)$. We find 
   \be \la{HamDef}
   p = { { d \phi \over d a }  a^4 \sqrt{V -{ 3 \over a^2 }}  \over \sqrt{ 3 - \half \left(a {d \phi \over d a }\right)^2 } }  ~,~~~~~~~~~~~ {\cal H } = p {d \phi \over d a } - \tilde I = 2 a^2 \sqrt{3} \sqrt{ V + \half { p^2 \over a^6} - { 3 \over a^2 } } 
   \ee 
  The Hamilton Jacobi equation is 
  \be 
   - \partial_a \tilde I = {\cal H } ( \phi, \partial_\phi \tilde I , a ) 
  \ee 
  and squaring each side we get
 \be \la{HJSq}
   ( \partial_a \tilde I)^2 =   
   12 a^4 V + { 6 \over a^2 } (\partial_\phi \tilde I)^2 - 36 a^2 
\ee 
We make now an ansatz 
 \be \la{Antt}
 \tilde I =   - 2 \left[A(\phi) a^3 - { 3 \over 2 }  B(\phi) a \right] + \dots  
 \ee 
 where the dots are subleading terms in $a$. 
  Inserting  this into \nref{HJSq} we get the equations 
   \be \la{FisHJ}
   V - 3 A^2 + 2 (A')^2 =0 ~,~~~~~~~~~~~ 0 = -1 + A B - 2 A' B'  
   \ee 

Let us give a more direct interpretation to $A$ and $B$. 
For that purpose let us first start with 
 the equation obeyed for $H(\phi)$ when we consider the zero curvature case. In other words, we can start from the equations 
\be \la{FlaSliE}
 3 H^2_f = \half \dot \phi^2_f + V ~,~~~~~~\dot H_f = - \half \dot \phi_f^2 
 \ee 
 which are equivalent to \nref{FriE} \nref{SFeq} in the case of zero spatial curvature. The subindex $f$ indicates that this a solution of the problem with flat slices.  We then think of $H_f$ as a function of the scalar field $\phi$ and write 
 \be 
 \partial_\phi H_f(\phi_f) = { \dot H_f \over \dot \phi_f } = - \half \dot \phi_f ~,~~~~~~~\to ~~~~~\dot \phi_f = - 2 \partial_\phi H_f 
 \ee 
 Substituting this into the first equation in  \nref{FlaSliE} we get 
 \be \la{Hfla}
 3 H^3_f = 2 (\partial_\phi H_f)^2 + V 
 \ee 
 This equation contains the same information as an arbitrary solution of \nref{FlaSliE}. We could solve it order by order in the slow roll factor, where the first approximation is $H_f \sim \sqrt{V/3}$. Since we are multiplying $H_f$ by a large $a$ factor in \nref{Antt}, in principle we need to consider all the higher order corrections in the slow roll factors. 
 Since \nref{Hfla} is the same as the equation that $A$ obeys in \nref{FisHJ} we can identify $A = H_f$.
 
 We should also note that, in general, for any solution we have that 
 \be \la{HExpI}
 H = - { 1 \over 6 } { 1 \over a^2 } \partial_a \tilde I 
 \ee 
 This simply follows from the fact that $H$ is the momentum conjugate to $a$. Notice that this says that $H = {\cal H }/(6a^2)$, with ${\cal H}$ is in \nref{HamDef}. 
 Then we see that saying that $A = H_f(\phi_r)$ is simply the first approximation to $H$ in \nref{HExpI}.  
  
To include the effects of spatial curvature, we   expand 
 \be \la{HexHf}
 H(\phi) = H_f(\phi) - \half  { 1 \over a^2 } b(\phi)
 \ee 

 We now expand the equations 
 \be \la{Fige}
 3 H^2  = { 3 \over a^2 } + \half \dot \phi^2 + V ~,~~~~~~\dot H  = { 1 \over a^2 } - \half \dot \phi^2 
 \ee
 to first order in $1/a^2$ using \nref{HexHf},    and also expand  the expression for the proper time as $t = t_f + \delta  $ so that 
  \be \la{Dfp}
 \dot \phi = \dot \phi_f (1 - \dot \delta  + \cdots ) 
 \ee 
 We find 
\be  -3  H_f b { 1 \over a^2} = - { 3 \over a^2 } - \dot \phi_f^2  \dot \delta ~,~~~~~~~~~
       - H_f' \dot \phi_f \dot \delta  -\half  { b' } \dot \phi { 1 \over a^2 }+  H_f b { 1 \over a^2 } = { 1 \over a^2 } + \dot \phi^2_f \dot \delta 
       \ee 
       Using the first to compute $\dot \delta $ and substituting in the second we get 
       \be \la{EqBb}
     0 =   2 b' H_f' - H_f b +1 
     \ee 
     This should be compared to the second equation in \nref{FisHJ} with 
     \be \la{GandH1}
     A(\phi) = H_f(\phi) ~,~~~~~~~~B(\phi)=   b(\phi) 
     \ee  

So we have derived the first deviation to the solution due to spatial curvature and, by computing $H$ \nref{HexHf} we have checked the expression \nref{Antt} via \nref{HExpI}. 
These are the first two terms in the action which we quoted in \nref{AsCt}. From \nref{EqBb} we get 
  $B \sim 1/H_f$ to leading order in the slow roll parameters, where we can neglect the first term in \nref{EqBb}. 

We are now ready to check the expression \nref{ActFi}. One might be tempted to add a term  $C(\phi)$ to \nref{Antt} and solve the equation to the next order in the $1/a$ expansion. This would give $C'=0$. This would be wrong, since we argued that $1/V_*$ does indeed have an $a$ dependence, see \nref{Shife}. The reason is that it is not so easy to disentangle the $\phi$ and $a$ dependence as we get to terms of order $\log a$.  Therefore we will check \nref{ActFi} using a different strategy as follows. 
We take the derivative of \nref{ActFi} to obtain a candidate expression for $H$  
\bea \la{Hfic}
H_r = - { 1 \over 6 } { 1 \over a^2_r } { \partial\tilde I \over \partial a_r } &=& A(\phi_r) - \half B(\phi_r)  { 1\over  a^2_r } + { i \over 3 a_r^3 } { a_r { \partial \over \partial a_r } { 3 \over V_* } } 
\cr 
&=&A(\phi_r) - \half B(\phi_r)  { 1\over  a^2_r } - { i \over 3 a_r^3 } {  2 \epsilon_* \over H_*^2}
\eea 
where we used that ${ a_r { \partial \over \partial a_r } { 3 \over V_* } }  = - { 2 \epsilon_* \over H_*^2 } $, a formula discussed around \nref{Shife}. 

We have already matched the first two terms to the leading expression for $H$ and its first correction due to the curvature. It remains to match the last correction. 
 
 The idea is that the last term in \nref{Hfic} should be matched against a correction    for $H$ that comes from a decaying perturbation that was generated around the region with the excursion into complex times. This decaying solution is present as terms going like $1/\alpha^3$  in both the scalar field and the metric \nref{vfLa} \nref{ResIm}. However, we need the form of these solution also for exponentially large $a$. We will find it as follows. First we will find the general form of the decaying solution in the region of exponentially large $a$. In this region we can find the solution by expanding \nref{Hfla} as $H_f \to H_{fsr} + \delta H$, where $H_{fsr}$ is the slow roll solution with the flat slicing and $\delta H$ is a perturbation that now obeys the equation 
 \be 
 6 H_{fsr} \delta H = 4 \partial_\phi H_{fsr} \partial_\phi \delta H 
 \ee 
 Solving this equation we find that 
 \be \la{SresC}
 \delta H \propto \exp\left( { 3 \over 2} \int d\phi { H_{fsr}  \over \partial_\phi H_{fsr} } \right) 
  \sim \exp \left( - 3 \int dt H \right) \propto { 1 \over a^3 } 
 \ee 
 This has the same form as the naive guess obtained from extrapolating the $1/\alpha^3$ term in \nref{ResIm} but it is now better justified. 
We now compute the change in the Hubble constant by using the solution   \nref{ResIm} which leads to 
\be \la{resbec}
H = { 1 \over \sqrt{ - g_{aa} } } { 1\over a^2 } \propto  H_* + \cdots  -   i { (2 \epsilon_*) \over H_*^2 }  { 2 \over 3 a^3 } 
\ee 
where the dots involve terms going like $1/a^2$ which we have already matched. 
One would be tempted to use this to  set the coefficient of \nref{SresC}. 
 However, in this solution we have a scalar field which also has a complex deviation  of the form  \nref{vfLa}  
 \be 
 \phi = \phi_* + \cdots + \delta \phi ~,~~~~~~~\delta \phi = - { i 2 \over 3 a^3 } { V'_* \over V_*}  { 1\over H_*^3}   
 \ee 
 where again the dots are real terms that are not relevant. 
 Expressing the first term  of \nref{resbec} in terms of the full $\phi$ we see we need to change 
 \be 
 H_* \to H_f(\phi) + \cdots - \partial_\phi H \delta \phi = H_f(\phi)  + \cdots  + i { 1 \over 3 a^3 } { 2 \epsilon_* \over H_*^2 } 
 \ee 
 
 So that now \nref{resbec} becomes  
 \be \la{HfiMa}
 H = H_f(\phi ) + \cdots -     { i\over 3 a^3 } { (2 \epsilon_*) \over H_*^2 }  
\ee 
where the dots are terms that go like $1/a^2$ that we have already matched. 

This can now be used to determine the coefficient of the $1/a^3$ solution in \nref{SresC}. We then see that the last term of \nref{Hfic} agrees with the last term in \nref{HfiMa}.

In conclusion,   the value of $H$ in our solution does indeed match the expression in \nref{Hfic} and this constitutes a test of the expression for the action \nref{ActFi}.

\subsection{Terms of order $\epsilon$ in the action }  
 \la{Epsilonterms}
 
 Here we simply want to give the explicit form of the integral of the terms of order $\epsilon$ in \nref{SecExpA}. 
 Writing the final action as 
 \be 
 iI = i I_0 +    4 \pi^2 { 3 \over V_*} \epsilon i \hat I_\epsilon 
 \ee 
 with $i I_0$ the expression in \nref{ActC} and  
\bea 
  i \hat I _\epsilon &=&  \alpha ^2-\frac{1}{6} (3 \pi +i) \left(\alpha ^2-1\right)^{3/2}-\frac{3}{2} (\pi +2 i)
   \sqrt{\alpha ^2-1}-\frac{2 i}{\sqrt{\alpha ^2-1}+i}+ 
   \cr 
   &~& + i \sqrt{\alpha ^2-1}
   \left(\alpha ^2+2\right) \log \left(1-i \sqrt{\alpha ^2-1}\right)+i \pi
   -\frac{11}{6}+\log 4 
 \eea 
For large $\alpha$ this becomes 
 \bea
  i \hat I _\epsilon  & = &   i \alpha ^3 (  \log \alpha -{1\over 6 })+\frac{1}{4} i \alpha  (6 \log  \alpha
    -11)+  (\log 4 - { 7 \over 2} + i \pi )  +o(1/\alpha)  
    \eea 
  Note that the terms of order $\alpha^0$ are just constant and there is no $\log \alpha $ term.

\section{Constraints from the Kontsevich-Segal-Witten criterion} 
 \la{KSW}
 
 The Kontsevich-Segal-Witten criterion \cite{Louko:1995jw,Kontsevich:2021dmb,Witten:2021nzp} is a proposed constraint on the type of allowed complex metrics as saddle points of the gravitational path integral. The idea is that we choose some real coordinates and have a complex metric. The allowed metrics are those that lead to well  actions for $p$-forms for all $p$. This turns out to imply the following. First we diagonalize the metric by real transformations so that it becomes diagonal $g_{ij } = \lambda_i \delta_{ij}$. Then the constraint is 
 \be 
  \la{KSWCon} 
  \sum_i | {\rm arg}(\lambda_i) | < \pi 
  \ee 
  The case of usual Lorentzian metrics is a limiting case where the argument of the time component approaches $\pi$. In that case it is clear that the other components of the metric do not have any leeway to get a small imaginary part. 
  Therefore one could expect that the no boundary geometries, which are complex  but approaching Lorentzian metrics, could be constrained by this criterion. In fact, \cite{Hertog:2023vot} found some interesting constaints, see also \cite{Jonas:2022uqb}.
   
  In this appendix, we explain the constraints found numerically in \cite{Hertog:2023vot} via our analytic approximations for the no boundary geometry. 
  
  \def\arg{{\rm arg}}
  
\begin{figure}[t]
    \begin{center}
     \includegraphics[scale=.7]{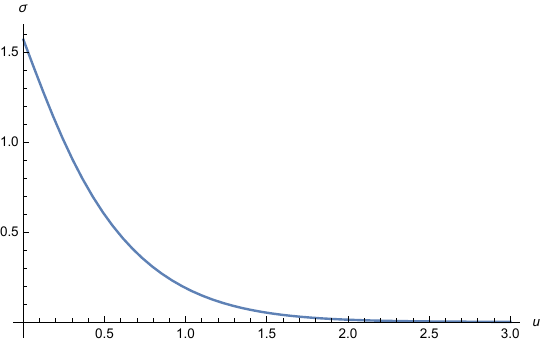} ~ ~~~~ \includegraphics[scale=.7]{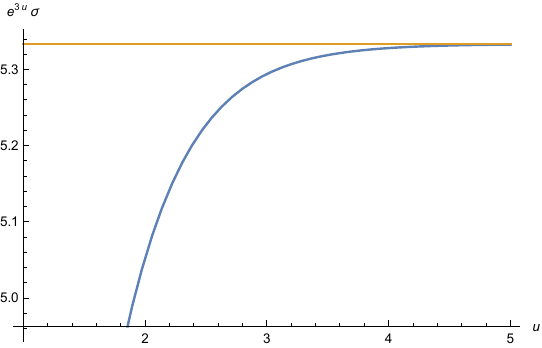}
    \end{center}
    ~~~~~~~~~~~~~~~~~~~~~~~~~~~~ ~ ~~~~~~~(a)~~~~~~~~~ ~~~~~~~~~~~~~~~~~~~~~~~~~~ ~~~~~~~~~~~~~~~(b)
     \caption{(a) Plot of the numerical result for $\sigma(u)$ solution of \nref{SigEq}. (b) We plot  $\sigma e^{3 u}$ to show that indeed $\sigma \sim \tilde c e^{ - 3 u } $ for large $u$. The horizontal line is $\tilde c = 5.333$.   }
     \label{SigmaFig}
 \end{figure}

  Let us start with the usual de Sitter metric \nref{dSGlob} and explore what the KSW criterion says in that case. 
  We are interested in finding a trajectory $\tau(u)$ such that the metric 
  \be 
  ds^2 =  e^{ i \pi } {\tau'(u)} ^2 + \cosh^2{\tau(u)} d\Omega_3^2 
  \ee 
  obeys the conditions stated above 
  \be \la{ExtConE}
   \left|\pi +  2 \arg[\tau'(u)] \right| + 6 \left|\arg [\cosh{\tau(u) } ]\right| \leq \pi 
   \ee 
   Parametrizing $\tau(u) = u + i \sigma(u)$, and looking for the extremal case where we replace the inequality by an equality we find the condition (assuming $\sigma>0$ and $\sigma' < 0$) 
   \be \la{SigEq} 
    0=\arg( 1 + i \sigma') +    {\arg [\cosh^3( u + i \sigma) ]}   ~,~~~~~~{\rm or } ~~{ d \sigma \over d u }  = - { y ( 3- y^2 ) \over (1 - 3 y^2) } ~,~~~~~ y =  \tanh u \tan \sigma(u)
\ee 
    In order to understand why there is a constraint we only need to expand this equation for small $\sigma$, which is what we expect to have at late times $u \gg 1$. Namely, we have 
    \be \la{orieq}
    \sigma' + 3 \sigma =0 ~,~~~~~~~\to~~~~~~ \sigma = \tilde c e^{ - 3 \tau } 
    \ee 
    where $\tilde c$ is some constant, which we expect to be an order one constant. In fact, numerically solving the equation, with the initial condition $\sigma(0) = \pi/2 $,  we get that the initial slope is down at an angle of $e^{ - i \pi/8}$ and then we find that 
    \be 
    \tilde c \sim 5.3 
    \ee 
    See figure \ref{SigmaFig}. 
    
    We now solve the equation analogous to \nref{ExtConE} but for the full no boundary geometry. For small values of $\tau$ the equation only gets small corrections of order $\epsilon$, which are not very important for $\tau $ of order one.  However, for large $\tau$ these corrections become important. In fact, 
   the solution that we found contains an imaginary  change in the scale factor \nref{Imava} which goes as 
    \be 
    a \to \cosh \tau (1 + i  \epsilon { 32 \over 9 }   e^{ - 3 \tau } ) ~,~~~~~  
        \ee 
        where we neglected irrelevant real changes. 
    The important point is that this correction   leads to an extra term at late times that modifies \nref{orieq} to 
    \be 
    \la{Newequ}
    \sigma' + 3 \sigma  +  { 32 \over 3}  \epsilon e^{ - 3 \tau } =0 
    \ee 
    whose solution is  
    \be \la{Newsso}
    \sigma = \tilde c e^{ - 3 \tau }  - {32 \over 3}   \epsilon \tau e^{ - 3 \tau } 
    \ee 
    where we picked a solution that for $\tau $ relatively small would match with \nref{orieq}. 
    
    Now the problem arises when \nref{Newsso} crosses the real axis. This will happen when 
    \be 
     \tau \epsilon  \sim { 3 \over 32   } \tilde c \sim 0.5 
     \ee 
    We can express this in terms of the tensor to scalar ratio $r =16 \epsilon$ and the number of e-folds, ${\cal N} \sim \tau $. Then we get the constraint 
    \be 
    { \cal N } r \leq 8 ~,~~~~~~~{\rm or } ~~~~~~r \leq   { 8 \over {\cal N } } \sim 0.13 
    \ee 
   where in the last expression we have set ${\cal N} =60$. 
   This roughly agrees with what was found for various potentials by a more detailed numerical analysis in figure 1 of 
    \cite{Hertog:2023vot}. We see here that the leading bound is on $\epsilon$. 
    
\section{Purely euclidean solution } 
\la{PESol}

In this appendix we consider the case where we set the inflaton at the reheating surface to be equal to $\phi_r$ and we let the size of the universe to adjust as it desires. In other words, we do not put any condition on $a_r$.  This implies that we have the condition that the extrinsic curvature should be zero. This means that the reheating surface is precise at the $S^3$ equator of the $S^4$. Now there is no period of Lorentzian evolution and the three sphere has the size of the horizon. 

We parametrize the scalar field as  
\be 
\phi = \phi_r + { V'_r \over V_r } \tilde \varphi 
\ee 
and the $S^4$ metric as $ds^2 = d \chi^2 + \cos^2 \chi d\Omega_3^2 $. Then  
we find 
\be 
\tilde \varphi = { 1 \over 1 + \sin \chi } -1  - \log( 1 + \sin \chi ) 
~,~~~~~~\chi \in [ 0 , {\pi \over 2} ] 
\ee 

\eject

\bibliographystyle{apsrev4-1long}
\bibliography{GeneralBibliography.bib}
\end{document}